# All-Optical Vector Measurement of Spin-Orbit-Induced Torques Using Both Polar and Quadratic Magneto-Optic Kerr Effects

Xin Fan[1,*], Alex R. Mellnik[2], Wenrui Wang[1,§], Neal Reynolds[2], Tao Wang[1], Halise Celik[1], Virginia O. Lorenz[1,§], Daniel C. Ralph[2,3], John Q. Xiao[1]

1. Department of Physics and Astronomy, University of Delaware, Newark, DE 19716 USA
2. Department of Physics, Cornell University, Ithaca, NY 14853 USA
3. Kavli Institute at Cornell, Ithaca, NY, 14853 USA

* Present address: Department of Physics and Astronomy, University of Denver, CO 80208 USA
§ Present address: Department of Physics, University of Illinois, Urbana, IL, 61801 USA

**Abstract**

We demonstrate that the magneto-optic-Kerr effect with normal light incidence can be used to obtain quantitative optical measurements of both components of spin-orbit-induced torque (both the antidamping and effective-field components) in heavy-metal/ferromagnet bilayers. This is achieved by analyzing the quadratic Kerr effect as well as the polar Kerr effect. The two effects can be distinguished by properly selecting the polarization of the incident light. We use this all-optical technique to determine the spin-orbit torques generated by a series of Pt/Permalloy samples, finding values in excellent agreement with spin-torque ferromagnetic resonance measurements.

Recent advances in the electrical control of magnetism [1-8] are exciting in part because they may lead to new technologies for nonvolatile magnetic memory and logic [9, 10]. Some of the mechanisms that are contenders to provide the highest-efficiency magnetic manipulation in practical device geometries involve current-induced torques arising from spin-orbit interactions, either in heavy-metal (HM)/ferromagnet (FM) bilayers [5, 6, 11, 12] or topological insulator (TI)/FM bilayers [13, 14]. Progress in this field, both for fundamental scientific understanding and practical applications, requires convenient, quantitative techniques for measuring the strength and direction of the spin-orbit torques, techniques that can be applied to a wide range of material systems. At present, the workhorse methods for measuring such torques are based on using magnetotransport signals for detecting magnetic reorientations in response to an applied current. For example, second-harmonic Hall effect measurements work well for measuring torques acting on a metallic magnetic layer with perpendicular magnetic anisotropy, but for magnets with in-plane anisotropy the need to separate out thermally-induced signals makes this technique more difficult to apply [11, 15, 16]. Spin-torque ferromagnetic resonance (ST-FMR) [6] can be used for metallic magnets with either perpendicular or in-plane anisotropy, but for very thin magnetic layers an artifact caused by spin pumping and the inverse spin Hall effect could in principle interfere with this method [17].

Here we demonstrate a simple all-optical technique for measuring current-induced torques that has a sensitivity comparable to the techniques based on magnetotransport detection, together with fewer artifacts and applicability to a very wide range of magnetic materials. The method is based on using the magneto-optic Kerr effect (MOKE) with normally-incident light to detect current-induced magnetic reorientation both perpendicular to the sample plane and in-plane, and therefore to measure both components of current-induced torque (there are two components because the magnitude of the magnetization remains fixed). Previously, some of us have shown that conventional polar MOKE with normal light incidence can be used to accurately measure the out-of-plane component of current-induced magnetic reorientation, and we suggested in that paper that longitudinal MOKE with oblique-angle light incidence might be used for measuring in-plane components [18]. However, longitudinal MOKE experiments are more challenging than polar MOKE, and the need for separate measurements with both normal and oblique-angle light incidence makes such experiments time-consuming and unattractive. Here we show that in-plane magnetic reorientations can actually be measured accurately and conveniently entirely with normal light incidence via a second-order (or quadratic) MOKE response. We demonstrate how polar and quadratic MOKE can be distinguished and separated by properly selecting the polarization of the incident light. We verify the accuracy of this method for measuring spin-orbit-induced torques by studying a series of Pt/Permalloy (Ni$_{81}$Fe$_{19}$ = Py) bilayers.

MOKE can be described as arising from a magnetization-dependent permittivity tensor, which can be expressed as a Taylor series in the components of the magnetization unit vector ***m*** [19]

$$\varepsilon_{ij}(\boldsymbol{m}) = \varepsilon_{ij}^{(0)} + \sum_k \varepsilon_{ijk}^{(1)} m_k + \sum_{k,l} \varepsilon_{ijkl}^{(2)} m_k m_l + ... \quad , \tag{1}$$

where $i, j, k, l = x, y, z$. When light interacts with a magnetic material, the light polarization will change depending on the magnetization orientation. The second term on the right side of Eq. (1) generates the first-order MOKE, that encompasses the well-known polar, longitudinal, and transverse MOKE responses [20]. The third term on the right in Eq. (1) leads to a second-order MOKE response, which is often referred to as quadratic MOKE [21]. This term is in general not negligible.

For the case of normally-incident light with linear polarization, the rotation of the polarization angle due to the magnetization can be written [22]

$$\psi(\boldsymbol{m}) = \alpha_{\text{Polar}} m_z + \beta_{\text{Quadratic}} m_x m_y + ... \quad , \tag{2}$$

where the $z$ direction is perpendicular to the magnetic film plane, the $x$ direction is parallel to the plane of the incident polarization, and $\alpha_{\text{Polar}}$ and $\beta_{\text{Quadratic}}$ are the coefficients for the polar MOKE and quadratic MOKE responses, respectively. One way to distinguish the polar MOKE and quadratic MOKE responses is by tuning the polarization of the light. If we define $\theta_M$ and $\phi_M$ as the polar and azimuthal angle of the magnetization, and $\phi_{\text{pol}}$ as the angle of the plane of polarization, then Eq. (2) can be rewritten (still assuming normally incident light with linear polarization) as

$$\psi(\boldsymbol{m}) = \alpha_{\text{Polar}} \cos\theta_M + \frac{1}{2}\beta_{\text{Quadratic}} \sin^2\theta_M \sin\left[2\left(\phi_M - \phi_{\text{pol}}\right)\right]. \tag{3}$$

As a result, the polar MOKE response does not depend on the polarization direction, while the quadratic MOKE depends on the polarization angle as $\propto \sin\left[2\left(\phi_M - \phi_{\text{pol}}\right)\right]$. Alternatively, if circularly-polarized incident light is used, the polar MOKE component yields no polarization change, while quadratic MOKE changes the polarization from circular to slightly elliptical [23]. Therefore, by controlling the polarization of the incident light, one can conveniently separate the two signals that are proportional to $m_z$ and $m_x m_y$, and thus measure current-induced magnetization rotation that results in changes to any of the magnetization components.

The existence of both polar and quadratic MOKE responses has a direct analog to the different processes that contribute to the Hall conductance in magnetic samples, due to the intimate connection between the permittivity tensor and the conductivity tensor in electromagnetism. Polar MOKE is analogous to the anomalous Hall effect ($\propto m_z$), while quadratic MOKE is the analog of the planar Hall effect ($\propto m_x m_y$).

We demonstrate MOKE-based spin-torque magnetometry using in-plane magnetized substrate/Pt(6 nm)/Permalloy($d_{Py}$) bilayers, with $d_{Py}$ ranging from 2 to 10 nm. The bilayers were grown at room temperature on C-axis epi-ready sapphire substrates in a magnetron sputtering system with a base pressure of $2 \times 10^{-9}$ Torr. After deposition of the Permalloy, 2 nm of Al was deposited and oxidized to form a protective barrier. Measurements of the electrical conductivity and the magnetization as a function of $d_{Py}$ are shown in ref. [23]. Based on the change in the bilayer conductivity as a function of $d_{Py}$, the conductivity of the Pt was estimated to be $3.2 \times 10^{6}$ $\Omega^{-1}$ m$^{-1}$ while the Py conductivity was approximately $3.6 \times 10^{6}$ $\Omega^{-1}$m$^{-1}$.

Our MOKE measurements of current-induced magnetization reorientation are conducted using the experimental geometry shown in Fig. 1. We apply an in-plane AC current, $I_{ac} \cos \omega t$, at 1013 Hz with $I_{ac}$ = 10 mA, and define the $x'$ axis as the direction of current flow, with $z' = z$ perpendicular to the sample plane. We initially align the magnetization along the $x'$ direction using an external field $H_{ext}$. The current-induced torque has components that rotate the magnetization locally within the sample plane (changing $\phi_M$) and perpendicular to the plane (changing $\theta_M$). The motion can be parameterized as resulting from two orthogonal effective-magnetic-field components $h_{y'}$ and $h_z$ as shown in Figure 1. To first order for an in-plane-magnetized sample (in MKS units)

$$\Delta \phi_M = \frac{h_{y'}}{H_{ext} + H_{anis}} , \quad (4)$$

$$\Delta \theta_M = \frac{h_z}{H_{ext} + H_{anis} + M_S - H_{anis\perp}} ,$$

where $H_{anis}$ is the in-plane anisotropy field (assumed to be in the $x'$ direction), $H_{anis\perp}$ is any out-of-plane anisotropy field due to interface or crystalline anisotropy, and $M_S$ is the saturation magnetization. For an ordinary transition-metal ferromagnet like Permalloy, the in-plane anisotropy is negligible and $M_S$ is much larger than any of the other field terms (for our sample of Pt (6nm)/Py (8nm), $\mu_0 M_s$ = 0.87 T as measured by vibrating sampling magnetometry, $\mu_0 H_{anis} = 0.23$ mT, and $\mu_0 H_{anis\perp} = 40$ mT as extracted from ferromagnetic resonance). This provides an additional method for separating the polar and quadratic MOKE signals for samples with in-plane magnetic anisotropy: for current-induced magnetic reorientations, the change in the polar MOKE signal ($\propto \Delta \theta_M$) should be approximately independent of applied field for $H_{ext} \ll M_S$, while the current-induced change in the quadratic MOKE signal should scale approximately as $1/H_{ext}$.

We analyze the Kerr rotations of the light polarization using the optical bridge apparatus shown in Fig. 1, with a mode-locked Ti:Sapphire laser working at 780 nm center wavelength. We use a half wave plate (labeled HWP-1) and a quarter wave plate (labeled QWP-1) to compensate a slight birefringence of the beam splitter and ensure that the light is initially linearly polarized along the *x'* axis when incident onto HWP-2 or QWP-2. To allow measurements in which linearly-polarized light is incident on the bilayer, and the polarization angle of the light can be adjusted relative to the sample magnetization, we rotate the light polarization by rotating the principle axis of a half wave plate (HWP-2) with respect to the *x'* axis. After the light is reflected from the sample and passes back through HWP-2, changes in the Kerr rotation angle are measured by using a polarizing beam splitter to separate the s- and p-components of the light and then analyzing the power difference by a balanced detector. Current-induced changes in the Kerr rotation, $\Delta\psi(\boldsymbol{m})$, are recorded by a lock-in amplifier locked to the frequency of the applied current. (The apparatus actually measures $-\Delta\psi(\boldsymbol{m})$ because the polarization rotation is reversed after the light passes through a half wave plate.) Since we study small current-induced rotations of the magnetic moment about an initial state with $\phi_M = 0$, $\theta_M = \pi/2$, by differentiation of Eq. (3) the expected change in the Kerr rotation signal is

$$-\Delta\psi(\boldsymbol{m}) = -\alpha_{\text{Polar}}\Delta\theta_M + \beta_{\text{Quadratic}}\cos 2\phi_{\text{pol}}\Delta\phi_M . \tag{5}$$

A derivation of this result using a Jones-matrix calculation is given in the supplementary material [23].)

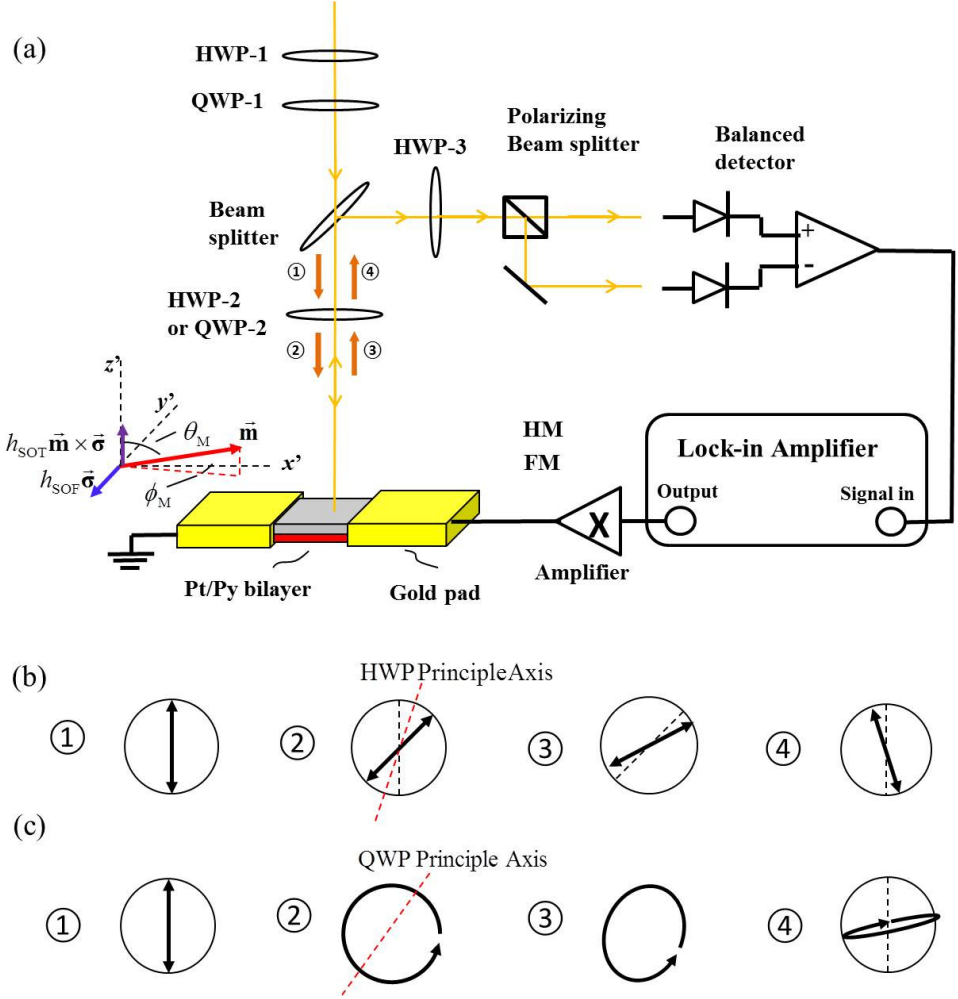

Figure 1 (a) Experimental setup for the optical detection of spin-orbit torques. For detecting the current-induced out-of-plane magnetization rotation we use a half wave plate HWP-2 before the sample. This is replaced by a quarter wave plate QWP-2 for detecting the current-induced in-plane magnetization rotation. (b) When the half wave plate HWP-2 is used, the light polarization at different points of the apparatus is as follows: ① The polarization is initially aligned in the $x'$ direction. ② Upon transmitting through the half wave plate HWP-2, the polarization is rotated by an angle of $\phi_{pol} = 2\phi_{HW}$, where $\phi_{HW}$ is the relative angle between the principle axis of the half wave plate and the $x'$ direction. ③ After the light is reflected from the magnetic material, the polarization changes to $2\phi_{HW} + \psi_{Kerr}(\phi_{pol} = 2\phi_{HW})$ due to polar and quadratic MOKE according to Eq. (5). ④ The polarization of the reflected beam is rotated to $-\psi_{Kerr}(\phi_{pol} = 2\phi_{HW})$ away from initial polarization after passing through the half wave plate HWP-2. (c) When the quarter wave plate QWP-2 is used instead, the light polarization at different points is as follows: ① the polarization is initially aligned in the $x'$ direction. ② After the quarter wave plate QWP-2, the polarization becomes circularly polarized. ③ Upon reflection from the magnetic material, the magnetization becomes elliptically polarized, due to the quadratic MOKE. ④ After passing through QWP-2 again, the polarization is rotated to the $y'$ direction with a perturbation due to the quadratic MOKE as described in Eq. (6).

In Fig. 2(a) we show the current-induced Kerr response as a function of swept magnetic field (in the $x'$ direction) for a sample with the layer structure wafer/Pt(6 nm)/Py(8 nm)/AlO$_x$ and different values of $\phi_{pol}$. For $\phi_{pol} = 45°$, where by Eq. (5) we expect the contribution from quadratic MOKE to be zero, we observe a simple step-like change in the current-induced Kerr signal near $H_{ext} = 0$, with the signal approximately independent of $H_{ext}$ on either side of the step. This is the behavior expected from the polar Kerr signal by itself, with the step near $H_{ext} = 0$ due to reversal of the magnetization, and with the weak magnetic-field dependence away from the step consistent with Eq. (4) for $H_{ext} \ll M_s$. As the polarization angle is rotated so that $\phi_{pol}$ differs from 45°, the form of the magnetic-field dependence of the current-induced Kerr signal changes dramatically, evolving from a simple step to the superposition of a step with an additional component that is approximately inversely proportional to $H_{ext}$. This is the signature of a significant quadratic MOKE signal in addition to the polar MOKE, with current-induced magnetization rotation within the sample plane providing the $1/H_{ext}$ dependence according to Eq. (4).

The current-induced torque in a Pt/Py bilayer contains contributions from both the Oersted field (with both an approximately uniform in-plane component together with out-of-plane components near the edges of the sample) and spin-orbit-induced torque. The spin-orbit torque can be described by an equivalent magnetic field that may also have components both in the sample plane (the "effective field" component) and out-of-plane (the "antidamping" component). The out-of-plane components due to the two mechanisms can be distinguished based on different symmetries with respect to reversing the magnetization; the Oersted field will not change upon magnetization reversal while the spin-orbit equivalent field should invert [18]. To isolate the out-of-plane Oersted field, we therefore plot the symmetric combination of polar Kerr signals, $\Delta\psi(+m_x) + \Delta\psi(-m_x)$, and to isolate the out-of-plane equivalent spin-orbit field, we plot the antisymmetric combination, $\Delta\psi(+m_x) - \Delta\psi(-m_x)$ (Fig. 2(b)), which is measured with linearly polarized light at $\phi_{pol} = 45°$. The procedure is the same as described in the supplementary material of Reference [18]. As expected, we find that the out-of-plane Oersted field is antisymmetric about the center of the wire, and the equivalent spin-orbit field is approximately constant across the wire width. Comparison of the measured out-of-plane Oersted field with a finite-element calculation of the Oersted field in a thin-film sample of finite width allows an accurate calibration of the equivalent spin-orbit field measured by the polar Kerr response.

The polar MOKE coefficient $\alpha_{Polar}$ is extracted from the calibration to be $(5.8 \pm 0.8) \times 10^{-3}$ [23], based on which we extract a value for the out-of-plane (i.e., antidamping) spin-orbit equivalent field of $\mu_0 h_{z,SO} = 0.068 \pm 0.010$ mT at a 10 mA current bias through the 50 μm strip. Using a simple parallel circuit model to account for the different resistivities of Pt and Py, we estimate that approximately 42% of the current flows through Pt, yielding a current density in the Pt of $j_{Pt} = 1.4 \times 10^{10}$ A/m$^2$. If we assume all of the antidamping-like torque is due to the spin Hall effect in the Pt layer, we determine a spin Hall angle $\theta_{Pt} = 0.082 \pm 0.012$ (using the formula $\theta_{SH} = (2e/\hbar)\mu_0 h_{z,SO} M_s d_{Py} / j_{Pt}$). This is consistent with the spin-Hall angle for Pt extracted using other torque-based measurement techniques [17, 24].

To more easily measure the in-plane component of the current-induced equivalent field, we substitute the half wave plate HWP-2 with a quarter wave plate QWP-2 with its principle axis set at 45° from the *x'*-axis to generate circularly polarized light incident upon the bilayer. Using Jones matrices to calculate the change in polarization due to each optical element, the expected current-induced change in the polarization is [23]

$$\Delta\psi(m) = -\beta_{Quadratic}\Delta\phi_M. \tag{6}$$

Therefore, only the quadratic MOKE effect should contribute a signal in this geometry, with no contribution from polar MOKE. Our measurements, shown in Fig.2(c), are consistent with the expectation that only $\Delta\phi_M$ (and not $\Delta\theta_M$) contributes to this signal, in that as a function of swept applied field $H_{ext}$ the step-like component seen in the polar-MOKE signal vanishes, leaving only a signal proportional to $1/H_{ext}$ away from $H_{ext} = 0$. We can calibrate the effective in-plane field produced by the current in the bilayer by using a metal strip fabricated on a printed circuit board attached to the back of the sample to apply a known oscillating in-plane external magnetic field, and measuring the quadratic MOKE signal due to this external field. From this calibration, the quadratic MOKE coefficient is found to be $\beta_{Quadratic} = (1.1 \pm 0.1) \times 10^{-4}$ [23]. For the Pt(6 nm)/Py(8 nm) sample with 10 mA current bias, the equivalent in-plane field produced by the current in the bilayer is $h_{y',SO} = 0.10 \pm 0.01$ mT. The uncertainty here arises mainly from inaccuracies in knowing the magnitude of the calibration field at the sample.

It is worth mentioning that the magnitude of $\alpha_{Polar}$ is almost two orders of magnitude greater than $\beta_{Quadratic}$, which is perhaps not surprising given that polar MOKE is a first-order process and quadratic MOKE is second-order. Nevertheless, because the out-of-plane magnetization reorientation $\Delta\theta_M$ is strongly suppressed by the demagnetization effect, the measured quadratic MOKE signal can still exceed the polar MOKE response in our thin-film bilayer samples.

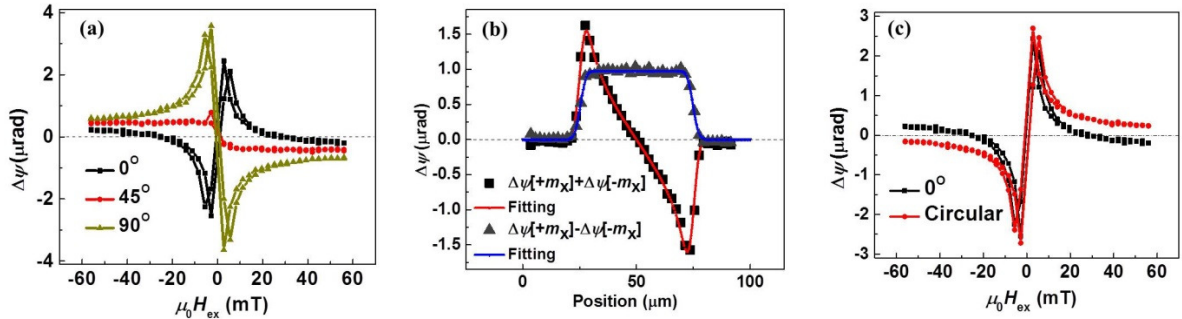

Figure 2 (a) Current-induced polar MOKE response with three different incident laser polarizations. (b) Separation of the out-of-plane field due to the Oersted field and the antidamping-like torque by spatial symmetry and dependence on the magnetization orientation. (c) Current-induced Kerr response with circularly polarized light and 0° linearly polarized light. The former only contains a quadratic MOKE response ($1/H_{ext}$-like) while the latter is a superposition of a quadratic MOKE response and a polar MOKE response (which has a step-like dependence on $H_{ext}$).

To further verify the accuracy of our MOKE-based spin-torque magnetometer, we measured samples with varying Py thickness: wafer/Pt(6 nm)/Py($d_{Py}$)/AlO$_x$, with $d_{Py}$ = 2-10 nm, and compared the results to ST-FMR performed on the same samples. We perform ST-FMR following the procedures described in ref. [14]: we apply a microwave current to the sample through a coplanar waveguide structure and detect a magnetic resonance signal via a rectified DC voltage. The magnitude of the symmetric part of the resonance allows a determination of the antidamping-like torque and the anti-symmetric part yields the in-plane effective-field component (for details, see Ref. [14]). The microwave current flowing through the sample is calibrated from a microwave reflection measurement; we do not assume as in Ref. [24] that the effective field component of the current-induced torque is due entirely to the Oersted field.

We plot in Fig. 3(a) the measured current-induced equivalent fields $h_{z,SO}$ (antidamping component) and $h_{y'}$ (effective-field component) determined by both MOKE and ST-FMR as a function of $d_{Py}$. These measured fields are normalized by the total surface current density ($I_{tot}/w$), where $w$ is the width of the sample. The two measurement techniques are in excellent quantitative agreement for both components. The strengths of both components of the equivalent field decrease as a function of increasing $d_{Py}$ in part because this corresponds to a decrease in the current density flowing in the Pt layer, however the dependences on $d_{Py}$ are different for the two components. This is as expected due to the physical differences between the antidamping spin Hall torque that acts at the interface of the magnetic layer and the in-plane Oersted field that acts throughout the thickness of the magnetic layer.

In Fig. 3(b) we take the measurements of the antidamping component from Fig. 3(a) and replot them in the form of a surface torque per unit area ($\tau_{AD,SO}/A = h_{z,SO}\mu_0 M_s d_{Py}$) normalized by the current density flowing just in the Pt layer, estimated from a simple parallel circuit model taking into account the different average resistivities of the Pt and Py layers. Over most of the range of Py thickness the torque is independent of $d_{Py}$, as expected for the surface torque due to the spin Hall effect arising from the Pt layer. The corresponding average spin Hall angle is $0.075 \pm 0.010$. There may be a small decrease in the strength of the torque for the 2 nm Pt layer, which is interesting in that it could hint at a decreased efficiency in the absorption of the incoming spin current for a very thin Py layer.

In Fig. 3(c) we replot the data for the in-plane $y'$ (effective field) component of the current-induced equivalent field taken from Fig. 3(a), but normalized versus the estimated current per unit lateral sample width flowing just in the Pt layer rather than the total current. For a pure Oersted field, the value should be 0.5, independent of Py thickness. We find that the measured equivalent field is indeed independent of $d_{Py}$, but the magnitude is somewhat larger than expected from a pure Oersted field. This discrepancy could be due to an inaccuracy in our simple parallel circuit model for estimating the current in the Pt (we neglect surface scattering, for example) or to the existence of a spin-orbit-induced effective field with an unexpected dependence on $d_{Py}$.

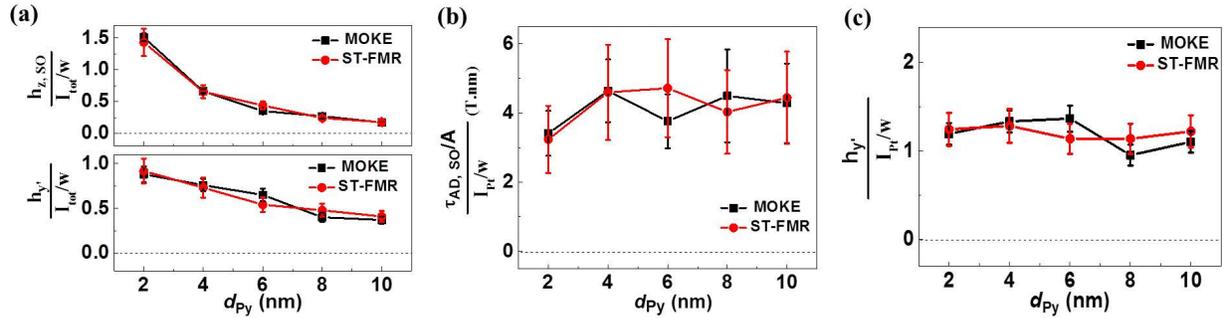

Figure 3 (a) The equivalent current-generated fields corresponding the antidamping-like component $h_{z,SO}$ and the in-plane effective-field-like component $h_{y'}$ normalized by the total current per unit lateral width in the bilayer. The uncertainties for the MOKE technique mostly arise from the fitting, while the uncertainties for the ST-FMR are mainly due to the determination of the microwave current. Excellent agreement is found between the MOKE and ST-FMR techniques. Lines are guides for eyes. (b) The antidamping torque $\tau_{AD,SO}/A$ normalized by the estimated current per unit sample width in the Pt layer. (c) and the in-plane equivalent field $h_{y'}$ normalized by the estimated current per unit sample width in the Pt layer.

In conclusion, we have demonstrated a convenient all-optical MOKE technique that can separately measure the antidamping-like and effective-field-like components of current-induced spin-orbit torque via polar MOKE and quadratic MOKE, respectively, with both measurements performed using normally-incident light. We find excellent agreement between the results of this technique and ST-FMR measurements for a series of Pt/Py bilayers with different Py thicknesses. We anticipate that MOKE magnetometry will be useful for rapid characterization of current-induced torques acting on a very wide range of materials.

**Acknowledgement**

Work at Delaware was supported by the Semiconductor Research Corporation through the Center for Nanoferroic Devices. Work at Cornell was supported by DARPA (N66001-11-1-4110) and the NSF (DMR-1010768). This research was performed in part at the Cornell NanoScale Facility, a node of the National Nanotechnology Infrastructure Network (NNIN), which is supported by the NSF (ECCS-0335765), and in the facilities of the NSF/MRSEC-funded Cornell Center for Materials Research (DMR-1120296).

# Supplementary Information for

# All-Optical Vector Measurement of Spin-Orbit-Induced Torques Using Both Polar and Quadratic Magneto-Optic Kerr Effects

Xin Fan[1,*], Alex R. Mellnik[2], Wenrui Wang[1, §], Neal Reynolds[2], Tao Wang[1], Halise Celik[1], Virginia O. Lorenz[1, §], Daniel C. Ralph[2,3], John Q. Xiao[1]

1. Department of Physics and Astronomy, University of Delaware, Newark, DE 19716 USA
2. Department of Physics, Cornell University, Ithaca, NY 14853 USA
3. Kavli Institute at Cornell, Ithaca, NY, 14853 USA

* Present address: Department of Physics and Astronomy, University of Denver, CO 80208 USA

§ Present address: Department of Physics, University of Illinois, Urbana, IL, 61801 USA

**Electrical and magnetic characterization of Py:**

We determine the electrical conductivity of our Py layers by comparing four-point resistance measurements of the Pt/Py bilayers to a control film with only 6 nm of Pt, shown in Fig. S1(a). The conductivity that we measure varies slightly as a function of Py thickness, with an average conductivity of about $3.6\times10^6$ $\Omega^{-1}$m$^{-1}$. We measure the saturation magnetization of our Py layers using vibrating sample magnetometry, shown in Fig. S1(b).

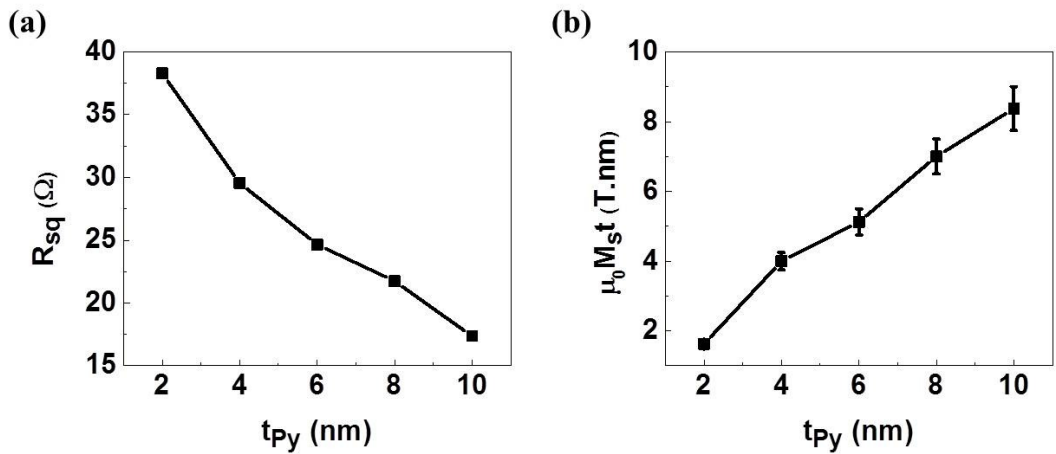

Figure S1 (a) Square resistance of the Pt/Py bilayers as a function of Py thickness. (b) $\mu_0 M_s t_{Py}$ as a function of Py thickness.

**Light with linear polarization incident on the bilayer:**

The total polarization rotation before and after the half wave plate HWP-2 can be derived using the method of Jones calculus, where polarization is described by a vector while transmission through wave plates and reflection from the magnetic samples are described by a matrix as shown in Table 1.

| | Initial Polarization | Half Wave Plate | Quarter Wave Plate | Magnetic sample |
|---|---|---|---|---|
| Jones Matrices /Vectors | $P_0 = \begin{bmatrix} 1 \\ 0 \end{bmatrix}$ | $M_{HW} = \begin{bmatrix} 1 & 0 \\ 0 & -1 \end{bmatrix}$ | $M_{QW} = \begin{bmatrix} 1 & 0 \\ 0 & -i \end{bmatrix}$ | $M_K = \xi \begin{bmatrix} 1+\frac{\beta_{Quadratic}}{2} & -\alpha_{Polar} m_z \\ \alpha_{Polar} m_z & 1-\frac{\beta_{Quadratic}}{2} \end{bmatrix}$ |

Table 1 List of Jones matrices/vectors used in this calculation. The initial polarization is set along the x'-axis. The Jones matrices $M$ in the table for half wave plate, quarter wave plate and magnetic sample [19, 20] are assuming the principle axis (fast axis of the wave plate or in-plane magnetization direction of the magnetic sample) is along the x'-axis. The matrices with arbitrary principle axis can be deduced as $R[\theta]MR[-\theta]$, where $R[\theta] = \begin{bmatrix} \cos\theta & -\sin\theta \\ \sin\theta & \cos\theta \end{bmatrix}$ and $\theta$ is the relative angle between the principle axis and the x'-axis. The factor $\xi$ in the Jones matrix for the magnetic sample captures the reflection loss, which does not affect the polarization change.

Therefore, the polarization at each stage in Fig. 1 of the main text can be calculated as

① $P_1 = \begin{bmatrix} 1 \\ 0 \end{bmatrix}$

② $P_2 = R[\phi_{HW}]M_{HW}R[-\phi_{HW}]P_1 = \begin{bmatrix} \cos 2\phi_{HW} \\ \sin 2\phi_{HW} \end{bmatrix}$

③
$P_3 = R[\phi_M]M_K R[-\phi_M]P_2 = \xi \left\{ \begin{bmatrix} \cos 2\phi_{HW} \\ \sin 2\phi_{HW} \end{bmatrix} + \alpha_{Polar} m_z \begin{bmatrix} -\sin 2\phi_{HW} \\ \cos 2\phi_{HW} \end{bmatrix} + \frac{\beta_{Quadratic}}{2}\begin{bmatrix} \cos(2\phi_M - 2\phi_{HW}) \\ \sin(2\phi_M - 2\phi_{HW}) \end{bmatrix} \right\}$

④
$P_4 = R[\phi_{HW}]M_{HW}R[-\phi_{HW}]P_3 = \xi \left\{ \begin{bmatrix} 1 \\ 0 \end{bmatrix} + \begin{bmatrix} \frac{1}{2}\beta_{Quadratic} \cos(4\phi_{HW} - 2\phi_M) \\ -\alpha_{Polar} m_z + \frac{1}{2}\beta_{Quadratic} \sin(4\phi_{HW} - 2\phi_M) \end{bmatrix} \right\},$

where $\phi_{HW}$ is the angle between the *x'*-axis and the principle axis of the half wave plate.

Therefore the total polarization angle rotation is $-\alpha_{Polar} m_z + \frac{1}{2}\beta_{Quadratic} \sin(4\phi_{HW} - 2\phi_M)$. By differentiating this polarization angle rotation near $\phi_M = 0$ and substituting $\phi_{HW} = \phi_{pol}/2$, we can derive Eq. (5) in the main text from $P_4$.

**Light with circular polarization incident on the bilayer:**

Using Jones calculus, the polarization at each stage in Fig. 1 of the main text where HWP-2 is replaced by QWP-2 can be calculated as

① $P_1 = \begin{bmatrix} 1 \\ 0 \end{bmatrix}$

② $P_2 = R[\pi/4] M_{QW} R[-\pi/4] P_1 = \dfrac{1-i}{2}\begin{bmatrix} 1 \\ i \end{bmatrix}$

③
$$P_3 = R[\phi_M] M_K R[-\phi_M] P_2 = \frac{1-i}{2}\xi\left\{(1-i\alpha_{Polar} m_z)\begin{bmatrix} 1 \\ i \end{bmatrix} + \frac{\beta_{Quadratic}(\cos\phi_M + i\sin\phi_M)^2}{2}\begin{bmatrix} 1 \\ -i \end{bmatrix}\right\}$$

④ $P_4 = R[\pi/4] M_{QW} R[-\pi/4] P_3 = \xi\left\{\begin{bmatrix} 0 \\ 1 \end{bmatrix} + \begin{bmatrix} \beta_{Quadratic} \dfrac{\sin 2\phi_M - i\cos 2\phi_M}{2} \\ i\alpha_{Polar} m_z \end{bmatrix}\right\}$

(S2)

Therefore the total polarization angle rotation is $\dfrac{\pi}{2} - \beta_{Quadratic}\dfrac{\sin 2\phi_M - i\cos 2\phi_M}{2}$. By differentiating this polarization rotation near $\phi_M = 0$, we can derive Eq. (6) in the main text.

**Extraction of Kerr rotation angle**

Initially the light is linearly polarized along the *x'* direction. For light with linear polarization incident on the bilayer, the reflected light after passing HWP-2 remains along the *x'*-axis with a slight deviation due to the MOKE as described by Eq. (S1). The principle axis of the analyzing wave plate HWP-3 is set to be 22.5° from the *x'*-axis. As a result, after passing through HWP-3, the light can be described by

$$R\left[\frac{\pi}{8}\right]M_{HW}R\left[-\frac{\pi}{8}\right]\xi\begin{bmatrix}1+\tfrac{1}{2}\beta_{Quadratic}\cos(4\phi_{HW}-2\phi_{M})\\-\alpha_{Polar}\cos\theta_{M}+\tfrac{1}{2}\beta_{Quadratic}\sin(4\phi_{HW}-2\phi_{M})\end{bmatrix}=$$

$$\frac{\xi}{\sqrt{2}}\begin{bmatrix}1+\tfrac{1}{2}\beta_{Quadratic}\cos(4\phi_{HW}-2\phi_{M})-\alpha_{Polar}\cos\theta_{M}+\tfrac{1}{2}\beta_{Quadratic}\sin(4\phi_{HW}-2\phi_{M})\\1+\tfrac{1}{2}\beta_{Quadratic}\cos(4\phi_{HW}-2\phi_{M})+\alpha_{Polar}\cos\theta_{M}-\tfrac{1}{2}\beta_{Quadratic}\sin(4\phi_{HW}-2\phi_{M})\end{bmatrix}=\begin{bmatrix}E_{x'}\\E_{y'}\end{bmatrix} \quad (S3)$$

After passing through the polarizing beam splitter, the light is split into two beams and analyzed by the balanced detector. The voltage output from the balanced detector is proportional to

$$|E_{x'}|^2-|E_{y'}|^2$$
$$=\xi^2[1+\tfrac{1}{2}\beta_{Quadratic}\cos(4\phi_{HW}-2\phi_{M})]\times[-\alpha_{Polar}\cos\theta_{M}+\tfrac{1}{2}\beta_{Quadratic}\sin(4\phi_{HW}-2\phi_{M})]$$
(S4)

By differentiating Eq. (S4), we determine the AC voltage output from the balanced detector to be $V_{Lock-in}=\xi^2\Delta\psi(m)$. On the other hand, when one of the inputs of the balanced detector is blocked, the DC component of the voltage output is $V_{DC}=\xi^2/2$. Therefore, the current-induced polarization rotation is extracted as $\Delta\psi(m)=\dfrac{V_{Lock-in}}{2V_{DC}}$.

Following the same process, it can be derived that the current-induced polarization rotation for light with circular polarization incident on the bilayer follows
$$\Delta\psi(m)=-\frac{V_{Lock-in}}{2V_{DC}}.$$

**Estimation of the MOKE coefficients**

Here we estimate the MOKE coefficients $\alpha_{Polar}$ and $\beta_{Quadratic}$. $\alpha_{Polar}$ is extracted from the polar MOKE data shown in Fig. 2(b) of the main text. The linescan $\Delta\psi(+m_x)+\Delta\psi(-m_x)$ is due to the out-of-plane Oersted field $h_{z,Oe}$ such that

$$\Delta\psi(+m_x)+\Delta\psi(-m_x)=\frac{2\alpha_{Polar}\langle h_{z,Oe}\rangle}{H_{ext}+H_{anis}+M_S-H_{anis\perp}},\text{where }\langle h_{z,Oe}\rangle\text{ is the average field in}$$

the region illuminated by the laser. The out-of-plane Oersted field can be calculated following Ampere's Law, $h_{z,Oe}=\dfrac{I}{2\pi w}\ln\dfrac{y'}{w-y'}$, where *w* is the width of the strip.

Through fitting the data as shown in Fig. 2 (b), we can extract $\alpha_{Polar}$ to be $(5.8 \pm 0.8) \times 10^{-3}$.

The MOKE signal measured with the calibration field $h_{y',Cal} = 0.08$ mT $\pm$ 0.008 mT applied along the y' direction is used to extract $\beta_{Quadratic}$. In this case, the magnetization will reorient in the x'-y' plane, $\Delta\phi_M = \dfrac{h_{y',Cal}}{H_{ext} + H_{anis}}$, following Eq. (4) in the main text. Hence the measured MOKE response can be deduced from Eq. (5) as $\Delta\psi_{Cal} = \beta_{Quadratic} \dfrac{h_{y',Cal}}{H_{ext} + H_{anis}}$. Using this expression, we fit the quadratic MOKE response measured under the calibration field, shown in Fig. S2, and obtain $\beta_{Quadratic} = (1.1 \pm 0.1) \times 10^{-4}$.

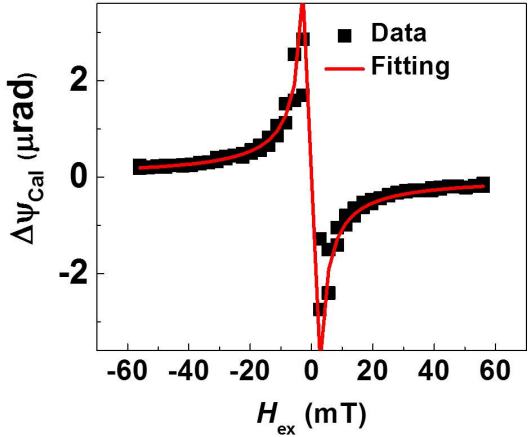

Figure S2. MOKE data when the calibration field is applied. The red curve is the fit using $-\Delta\psi(\mathbf{m}) = \beta_{Quadratic} \dfrac{h_{y',Cal}}{H_{ext} + H_{anis}}$.

**Laser polarization angle dependent MOKE response**

We have performed a laser-polarization-angle-dependent MOKE study to verify the angular dependence of the Kerr coefficients assumed in Eq. (5) in the main text. Within linear response, the current-induced Kerr rotation in general should be described as

$$\Delta\psi(\boldsymbol{m}) = a(\phi_{pol})\Delta\theta_M + b(\phi_{pol})\Delta\phi_M, \qquad (S5)$$

where $a(\phi_{pol})$ and $b(\phi_{pol})$ are the MOKE coefficients that may depend on the polarization angle while $\Delta\theta_M$ and $\Delta\phi_M$ are the current-induced polar and azimuthal angle change, which are independent of the polarization. Using Eq. (4) and the fields $h_z$ and $h_{y'}$ derived in the main text when passing 10 mA current through the 50 μm sample strip, we extract $\Delta\theta_M = 77\,\mu\text{rad} \pm 11\,\mu\text{rad}$ and $\Delta\phi_M = (1.1 \pm 0.1)\times 10^{-4}\,\frac{0.1\text{mT}}{\mu_0 H_{ex}}$. Therefore, $a(\phi_{pol})$ and $b(\phi_{pol})$ can be extracted from the MOKE data measured at different polarizations through linear regression. The extracted data, shown in Fig. S3, reveals that indeed $a(\phi_{pol})$ is nearly independent with polarization and $b(\phi_{pol})$ has a cosine dependence on the polarization, which confirms Eq. (5) in the main text.

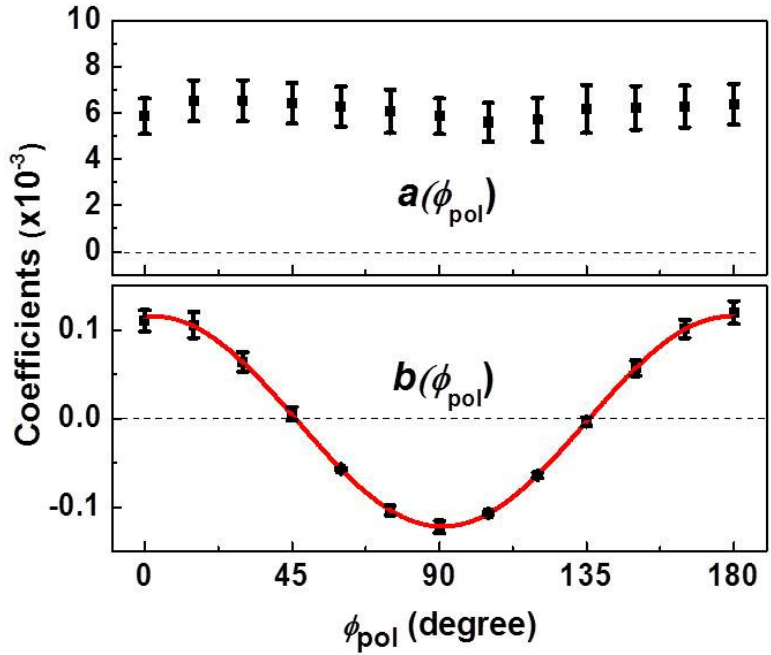

Figure S3. MOKE coefficients plotted as a function of laser polarization. The red curve in the bottom panel is a sinusoidal fit to $\cos 2\phi_{pol}$.